\documentstyle[twoside,fleqn,espcrc2,epsfig]{article}

\newcommand{\be}{\begin{equation}}
\newcommand{\ee}{\end{equation}}

\def \gta {\mathrel{\vcenter{\hbox{$>$}\nointerlineskip\hbox{$\sim$}}}}

\title{Leptogenesis via preheating \thanks{Talk given at Valencia '99, based on a paper by G.F.\ Giudice, M.\ Peloso, A.\ Riotto and I.\ Tkachev \cite{gprt}.}}

\author{Marco Peloso \address{Scuola Internazionale Superiore di
Studi Avanzati \\ Via Beirut 4, 34014 Trieste, Italy \\ and \\
Istituto Nazionale di Fisica Nucleare, Sez.\ di Trieste \\ Via
Valerio 2, I-34127 Trieste, Italy}}

\begin{document}

\begin{abstract}
Leptogenesis constitues a very simple scenario to achieve the
baryon asymmetry that we observe today. It requires only the
presence of right handed neutrinos (which arise very naturally
in many extensions of the Standard Model) and depends crucially
on the mechanism responsible for their production. In
particular, when their mass exceeds the inflaton mass
($10^{13}\,$GeV in chaotic inflation) only non perturbative
production can occur. It is shown here that non perturbative
production of fermions in an expanding Universe is a very
efficient mechanism up to masses of order of $\left( 10^{17} -
10^{18} \right) \,$GeV, and that it can be easily applied to
solve the baryon asymmetry problem.
\end{abstract}

\maketitle

\section{Introduction}

Inflation is at present the most natural and accepted solution to many
cosmological puzzles, such as the flatness, the horizon, and the monopole
problems. However, the Universe after inflation looks very different from
the one we observe today, since the huge expansion leaves it empty of
radiation and matter.

The only form of energy which is not diluted away after this
phase is the vacuum energy of the field (or fields) which leads
inflation itself, namely the inflaton field. The process of
converting this energy into the radiation and matter which we
see around has been named {\it reheating} and it has been the
object of intense studies in the past decades.

Apart from this energy conversion, the processes which govern the first
instant of the Universe must also favor the production of matter over
antimatter. Indeed, on one hand direct observations put strong limits to
the presence of antimatter in our cluster, while on the other
considerations about how the light elements abundances were formed when
the Universe was about $1 \:\mbox{MeV}$ hot lead us to conclude that the
difference between the number density of baryons and that of antibaryons
is about $10^{-\,10}$ if normalized to the entropy density of the
Universe.

Until now, many mechanisms for the generation of the baryon
($B$) asymmetry have been proposed (for a recent review, see
\cite{rt}).

Grand Unified Theories (GUTs) unify the strong and the
electroweak interactions and predict baryon number violation at
the tree level. They are --- therefore --- perfect candidates
for a theory of baryogenesis. There, the out-of-equilibrium
decay of superheavy particles can explain the observed baryon
asymmetry.

Another plausible scenario is the theory of electroweak
baryogenesis (in the context of the Minimal Supersymmetric
Standard Model), where baryon number violations take place at
the quantum level due to the usoppressed and baryon number
violating sphalerons in the hot plasma.

Sphaleron transitions leave unchanged the combination $B - L$
(where $L$ is the lepton number) and play the role of converting
an initial lepton asymmetry to a final mixture of nonvanishing
$B$ and $L\,$. This is the key idea of the theories of
baryogenesis via leptogenesis. Once a lepton asymmetry is
produced, sphaleronic transitions convert a fraction of it into
baryon number. In the Standard Model the following relation for
the final amount of the baryon asymmetry holds:
\be
B=\left( \frac{8 n_f + 4 n_H}{22 n_f + 13 n_H}\right) \: \left(
B - L \right) \;,
\ee
where $n_H$ is the number of Higgs doublets and $n_f$ is the
number of fermion generations.

Adding right-handed Majorana neutrinos to the SM breaks $B-L\,$, and the
primordial lepton asymmetry may be generated by their out-of-equilibrium
decay. This simple extension of the SM can be embedded into GUTs with
gauge groups containing $SO(10)$. Heavy right-handed Majorana neutrinos
are particularly welcome to explain the smallness of the light neutrino
masses via the see-saw mechanism \cite{gellm}.  In most of the models
proposed so far to generate neutrino masses, the latter are of the
Majorana type, which implies the existence of interactions which violate
the lepton number and thus render leptogenesis particularly attractive.

The leptogenesis scenario depends crucially on the details of the
reheating process after inflation and on the production mechanism that was
responsible for populating the Universe with right-handed neutrinos.

The simplest way to envision this reheating process is if the comoving
energy density in the zero mode of the inflaton decays {\it
perturbatively} into ordinary particles, which then scatter and thermalize
to form a thermal background \cite{dolg,abbot}. Of particular interest is
a quantity known as the reheat temperature, which is given by
\be
T_{RH}= \sqrt{ \Gamma_\phi M_p } \;,
\ee
where $\Gamma_\phi$ is the inflaton decay rate.

In order to produce a sufficient amount of heavy right-handed
neutrinos with mass $M_N\:$, it is commonly assumed that
$T_{RH}\gta M_N\;$.

However, $T_{RH}$ is constrained by the so called gravitino
problem \cite{gravi} to be less than about $10^{10}\:$GeV.

This bound increases if the heavy neutrinos are produced
directly through the inflaton decay process. This would be
kinematically accessible only if \footnote{The last equality is required by the observed density and temperature fluctuations.} 
\be
M_{N}< M_\phi \simeq 10^{13}\: {\rm GeV}\;.
\ee

The outlook for leptogenesis might be brightened even further 
with the realization that reheating may differ significantly
from the simple picture described above \cite{explo,KT}. In the
first stage of reheating (namely in the first dozen or so
oscillations), called {\it preheating} \cite{explo}, nonlinear
quantum effects may lead to extremely effective dissipative
dynamics and explosive particle production, even when single
particle decay is kinematically forbidden.

It has been already shown that preheating plays an extremely important
role in baryogenesis, in particular for the production of superheavy GUT
bosons \cite{klr,krt}. The application of preheating that will be instead
presented here is a very new one \cite{gprt} and it consists in enriching
the leptogenesis scenario with the possibility of producing right-handed
neutrinos which are too heavy to be generated in the usual perturbative
ways.

In what follows it will be shown (both with numerical
calculations and with some analytical estimates \footnote{New
analytical expressions with respect to the paper \cite{gprt} ---
where the reader is addressed for a more general treatment of
leptogenesis --- will be presented here.}) that the production
of a generic superheavy fermionic field during preheating is a
very efficient mechanism up to masses of order of $10^{18}
\:$GeV. As a consequence, it will be pointed out that a
sufficient $L$ asymmetry can be generated if the fermionic field
is a right-handed neutrino.

\section{Production of fermions during preheating}

The coupling inflaton --- fermions gives to the latter the effective time dependent mass
\be \label{mass1}
m \left( t \right) = m_X + g \, \phi \left( t \right) \;,
\ee
where $m_X$ is the bare fermionic mass.

This changing mass induces the creation of $X$ particles. To see
this, one has to diagonalize the hamiltonian associated to the
Dirac equation
\be
\left( \frac{i}{a} \gamma^\mu \partial_\mu + i  \frac{3}{2} H \gamma^0-m
\right) X = 0
\ee
that is written here for an expanding flat Universe ($H$ is the Hubble constant) and in conformal time $\eta\;$.

Once the diagonalization is performed (see \cite{gprt} for details) the number of produced fermions is given by
\begin{eqnarray} \label{n}
n \left( \eta \right) &\!\!\!\!\!=\!\!\!\!\!& \frac{1}{\pi^2 \, a^3 \left( \eta \right)}\, \int_0^{\infty} \,d\,k \: k^2 \: n_k^2 \left( \eta \right) \nonumber \\
n_k^2 \left( \eta \right) &\!\!\!\!\!=\!\!\!\!\!& \frac{\omega - 2 k \, \mbox{Re} \left( u_+^* u_- \right)\!-a\,m \left( 1 - 2 u_+^* u_+ \right)}{2\:\omega}
\end{eqnarray}
where $\omega^2 = k^2 +m \left(t \right)^2 a^2\,$, $k$ is the
momentum of the produced particle, and $a$ is the scale factor
of the Universe. The functions $u_\pm$ are, respectively, the
positive and negative eigenfunctions of $\gamma_0$ which enter
in the decomposition of the fermionic field. They satisfy the
uncoupled equations
\be
\left[ \frac{d^2}{d\eta^2} +\omega^2 \pm i (a'm+am')\right]u_\pm(k)=0\;,
\ee
with initial conditions at time $\eta = 0$ given by 
\footnote{Note that with these values $n \left( 0 \right) =
0\:$.}
\begin{eqnarray}
u_\pm (0) &\!\!=\!\!& \sqrt{\frac{1}{2} \left( 1 \mp \frac{ma}{\omega} \right)} \nonumber\\
u^{\prime}_\pm (0) &\!\!=\!\!& i \, ku_\mp (0) \mp i\, a \,mu_\pm(0)\;.
\end{eqnarray}

A measure of the strength of the coupling inflaton --- fermions
is given by the parameter
\be
q \equiv g^2 \, \phi^2 \left( 0 \right) / 4 \, m_\phi^2\;,
\ee
where $\phi \left( 0 \right) \simeq 0.28 M_p$ is the value of
the inflaton at the end of the inflation and $m_\phi \simeq
10^{13}\:$GeV is the inflaton mass.

\begin{figure}[htd]
\vspace{9pt}
\epsfig{file=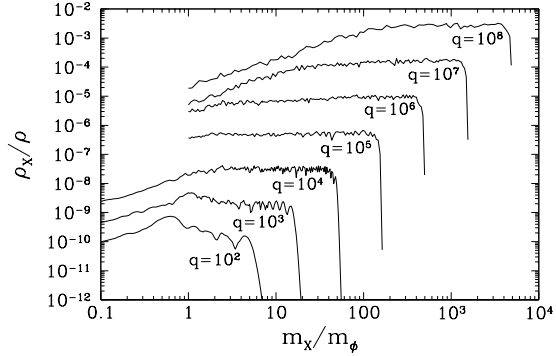,width= \linewidth}
\caption{Ratio between the energy density of the produced fermions and of the inflaton field, for many values of the fermion mass and of $q\,$.}
\label{scan}
\end{figure}

In Fig.\ref{scan} the results of a numerical integration of the
equations given above are reported. For a wide range of $m_X$
and $q$ the value of the ratio $\rho_X / \rho$ is shown, where
$\,\rho_X = m_X \cdot n_X\,$ is the energy density transferred
to the fermions and $\rho$ is the initial energy density stored
in the inflaton field.

One can notice the presence of the cut-off $\,m_{\mbox{max}}
\sim q^{1/2} / 2\,$ in the fermionic mass. For $m_X$ higher than
this value the production is substantially zero, while for lower
values it increases linearly with $q$ and is independent from
$m_X\:$. This last statement seems not to be completely
confirmed by Fig.\ref{scan}, where the value $\rho_X$ increases
slowly with $m_X\:$. However, as the analytical estimate below
will confirm, the scaling stated here is exact, while the slope
that one sees in Fig.\ref{scan} for high $q$ and small $m_X$ is
a numerical artifact. The problem is that --- due to limited
computational resources --- $\rho_X$ reported in Fig.\ref{scan}
is evaluated after $20$ oscillations of the inflaton field.
However, the number of oscillations of $\phi$ that one needs for
$\rho_X$ to saturate increases with $q\,$, and a slope in
$\rho_X$ appears for those values for which the saturation has
not yet taken place.

\begin{figure}[htd]
\begin{center} 
\vspace{9pt}
\epsfig{file=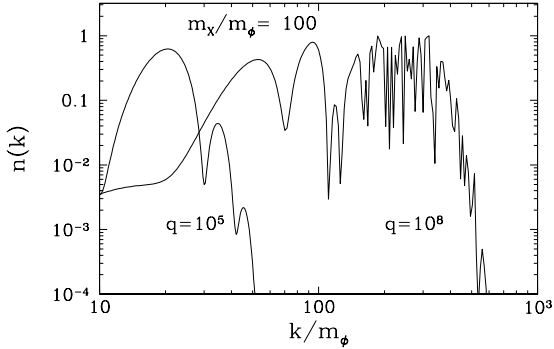,width= \linewidth}
\caption{Spectrum in momenta of the produced particles for a given fermion mass and two different values of $q\,$.}
\label{spect}
\end{center}
\end{figure}

To go on, it is important to understand the behavior of
$n_k\,$, that is the number density of fermions produced with a
given momentum $k\,$. All these spectra exhibit the qualitative
behavior that is reported in Fig.\ref{spect} for $m_X = 100 \,
m_\phi$ and for two values of $q\,$.

Also here one can find a threshold $\,k_{\mbox{max}}\,$ in the
production. For $k <  k_{\mbox{max}}\;\,,\,\; n_k$ approximately
saturates to $1$ (filling the whole allowed Fermi sphere), while
for momenta bigger than $k_{\mbox{max}}$ the production does not
occur. Thus one can estimate [see Eq.~(\ref{n})]
\be \label{rho}
\rho_X = m_X \cdot n_X \propto m_X \cdot k_{\mbox{max}}^3 \;\;.
\ee

As it is shown in Fig.\ref{spect}, $k_{\mbox{max}}$ increases
with $q\,$. More precisely, the analytical estimate that will
be given below indicates $k_{\mbox{max}} \propto \left( q / m_X
\right)^{1/3}$. From this value and from Eq.~(\ref{rho})
one immediately gets the scaling $\rho_X \propto q$ stated above.

To understand the scaling of $k_{\mbox{max}}$, one can
differentiate Eq.~(\ref{n}):
\be \label{dndeta}
\frac{d\,n_k}{d\,\eta} = - \frac{1}{\omega^2} \: \frac{d\,\omega}{d\,\eta} \:\frac{d^2\,\vert u_+ \vert^2 / d\,\eta^2}{8\,\left( m\,a\right)}\;.
\ee
This expression suggests that $n_k$ changes significantly when
the total mass $m$ approaches zero and when the variation of the
frequency $\omega$ is non-negligible. Indeed, our numerical
results indicate that the production of fermions occurs for very
short intervals about the zeros of $m\,$. The estimate that is
here presented consists in neglecting the last factor in
Eq.~(\ref{dndeta}) and, as it is customary done for preheating
of bosons, in assuming that the production occurs only for
$\left| \omega' / \omega^2  \right| > 1  \;,$ that is in a
regime of non adiabatic change of the frequency \footnote{The
results that will be presented here can be obtained (in a more
lengthy way) also without this assumptions and hence do not
depend on the approximations made here.}.

In terms of the number $N$ of oscillations of the inflaton field, the
non adiabaticity condition rewrites
\be \label{nonad}
\frac{k^2}{a^2} \, < \, \left( \frac{m}{2\,\pi}\frac{dm}{dN} \right)^{2/3}
- m^2  \; ,
\ee
where masses and momenta are given in units of the inflaton mass.

The cut-off in the production for $m_X > m_{\mbox{max}}$ can be now understood by rewriting Eq.~(\ref{mass1}) in terms of $N\,$:
\be \label{mass2}
m = m_X + 2q^{1/2}\frac{\phi}{\phi(0)}
\approx m_X + \frac{q^{1/2}}{\pi N} \,\mbox{cos}\, 2\,\pi\,N \,.
\ee
At large $m_X$ the first zero of $m$ can be potentially reached
at $N = 1/2$. However, for $m_X$ greater than $2q^{1/2} / \pi$,
the total mass $m$ never vanishes and Eq. (\ref{nonad}) is never
satisfied  \footnote{While numerical coefficient based on the
second approximate equality in the r.h.s of Eq. (\ref{mass2})
should not being taken seriously, numerical integration shows
that $2{\phi}/{\phi(0)} \approx - 0.5$ in the point of first
minimum of $\phi$, which gives the cut-off in the proper place,
$m_X \approx q^{1/2} / 2$.}.

For $m_X < m_{\mbox{max}}\,$, the total mass vanishes at the points $N_0$ given by ($N_{\mbox{max}}$ is the greatest $N_0\,$)
\be \label{zeros}
N_0 = -  \frac{q^{1/2}}{\pi \: m_X} \,\mbox{cos}\, 2\,\pi\,N_0 \;\;,\;\; N_{max} \simeq \frac{q^{1/2}}{\pi \: m_X} \;.
\ee

To find $k_{\mbox{max}}\,$, one has to maximize r.h.s. of Eq.~(\ref{nonad}). The maximum occurs when the two terms in r.h.s. are of the same order. Thus, a good estimate for $k_{\mbox{max}}$ can be achieved by solving
\be
\left( \frac{m\,m'}{2\,\pi} \right)^{2/3} \simeq m^2 \simeq \left( \frac{k}{a} \right)^2
\ee
expanding $m$ and $m'$ for $N$ close to $N_0\,$. Doing so, one
gets
\be
k \simeq a\,m \sim \vert \mbox{cos}\, 2\,\pi\,N_0 \vert^{1/3} \, q^{1/6} \, N_0^{1/3}.
\ee

Since $k$ grows with $N_0\;,\, k_{max}$ has to be evaluated in
$N_{max}\;$. Thus the particles with the highest momentum are generated in the last zero $N_{max}$ of $m\,$. This gives the stated result 
\be
k_{max} \propto \left( \frac{q}{M} \right)^{1/3} \;.
\ee

\section{Baryogenesis}

The previous section shows that preheating can produce fermions with the 
energy density
\be \label{result}
\left( \frac{\rho_X}{\rho_\phi} \right) \sim 10^{-\,11} \: q \;\;\mbox{up
to}\;\; M_X = \frac{q^{1/2}}{2} \: m_\phi \;.
\ee

This result can be useful for leptogenesis if one identifies the
$X$ particles with right-handed neutrinos $N$ and if these heavy
neutrinos decay in a $CP$ violating manner into left handed neutrinos plus higgs particles.

Eq.~(\ref{result}) is obtained with the tacitly assumption that
the produced fermions are stable. However this estimate does not
apply if the right handed neutrinos have a decay lifetime larger
than the typical time-scale of the inflaton oscillation
$m_\phi^{-1}\;$.

Another requirement that must be put is that the right handed
neutrinos do not annihilate into inflaton quanta more rapidly
than they decay in left handed neutrinos, otherwise the produced
lepton asymmetry would be negligible. 

Both the above requirements can be typically satisfied
\footnote{One must also ask that lepton number violating
scatterings do not deplete the lepton asymmetry generated bye
the decay of the right handed neutrinos. See \cite{gprt} for
details.} for masses $M_N \leq 10^{15} \:$GeV and for couplings $q
\leq 10^{10}\:$. 

Once the right handed neutrinos decay and once the lepton
asymmetry produced by the decay is converted in baryon asymmetry
by sphaleron transitions, Eq.~(\ref{result}) gives  \cite{gprt}
\begin{eqnarray} \label{bar}
&& B \:\equiv\: \frac{n_B}{s} \:= \nonumber \\
3 \cdot 10^{-\,7} \!\!\!\!\!\!&\varepsilon&\!\!\!\!\!\! \left(
\frac{T_{rh}}{10^{10} \:\mbox{GeV}} \right) \left( \frac{10^{15}
\:\mbox{GeV}}{M_N} \right) \left( \frac{q}{10^{10}} \right),
\end{eqnarray}
where $T_{rh}$ is the reheating temperature and $\varepsilon$ is
the amount of $CP$ violation in the decay of the heavy neutrinos
\footnote{Eq~(\ref{bar}) holds if the energy density of the
right handed neutrinos never dominates over the thermal one.
However, if the neutrinos have a very late decay they can
eventually dominate over the thermal bath, since their energy
does not redshift with the expansion of the Universe. If it is
the case, the decay of the neutrinos gives both the baryon
number and the entropy of the thermal bath, and the resulting
baryon asymmetry is given by
\be
B = 8\cdot 10^{-\,3} \left( \frac{m}{10^{-\,6} \: \mbox{eV}} \right)^{1/2}
\:\varepsilon \;\;,
\ee
where $m$ is the mass of the light neutrinos.}.

Eq~(\ref{bar}) reproduces for very natural values of the
parameters the observed baryon asymmetry of the Universe. This
result leads thus to conclude that baryogenesis via leptogenesis
is a viable option also if one considers right-handed neutrinos
with mass higher than the inflaton one.

\section{Acknowledgements}

I would like to thank G.F.\ Giudice, A.\ Riotto, and I.\ Tkachev, with
whom the results presented in this talk were obtained. I'm also grateful
to the organizators of the Conference for the very friendly and
stimulating atmosphere.  This work is partially supported by the EEC TMR
network ``Beyond the Standard Model'', contract no.\ FMRX-CT96-0090.

\end{document}